\documentclass[12pt]{article}

\def\bard{d\kern-2.5pt\raise 3pt\hbox{-}}
 
\title {Chiral Shielding}
\medskip
\author{L.~Babukhadia$^{a}$, Ya.~A.~Berdnikov$^{b}$, A.~N.~Ivanov$^{b,c}$ \\ and M.~D.~Scadron$^{a}$ \\ 
	$^{a}$ Physics Dept., Univ. of Arizona, \\ Tucson, AZ 85721 USA \\
	$^{b}$ Nuclear Physics Dept., State Technical Univ., \\ 195251 St.~Petersburg, 
	Russian Federation \\
	$^{c}$ present address, Inst. f{\"u}r Kernphysik, Technische Univ., \\ A--$1040$, Wien, Austria}

\begin{document}
\maketitle

\begin{abstract}

We demonstrate how a chiral soft pion theorem (SPT) shields the scalar 
meson ground state isoscalar $\sigma(600-700)$ and isospinor $\kappa(800-900)$ 
from detection in $a_{1}\rightarrow\pi(\pi\pi)_{\mathrm{swave}}$,
$\gamma\gamma\rightarrow2\pi^{0}$, $\pi^{-}p\rightarrow\pi^{-}\pi^{+}n$ and
$K^{-}p\rightarrow K^{-}\pi^{+}n$ processes.
While pseudoscalar meson PVV transitions are known to be determined by (only)
quark loop diagrams, the above SPT also constrains scalar meson SVV transitions to be
governed (only) by meson loop diagrams. 
We apply this latter SVV theorem to $a_{0}\rightarrow\gamma\gamma$ and
$f_{0}\rightarrow\gamma\gamma$ decays.

\noindent
pacs: 11.30.Rd, 12.39.-x, 12.39.Ki

\end{abstract}

\baselineskip 22pt

\section{Introduction}

The recent plethora of scalar meson papers appearing in the Los Alamos
archives~\cite{plethora} stresses once again the importance but difficulty 
in observing 
the ground state $I=0$ and $I=1/2$ scalar mesons $\sigma(600-700)$ and 
$\kappa(800-900)$. 
Although these resonances were first listed in many of the $1960$--$70$
particle data group
(PDG) tables, they were later removed in the mid $1970$'s in favor of the 
higher mass $\epsilon(1300)$ and $\kappa(1400)$.
Chiral symmetry shields the $\sigma(600-700)$ and $\kappa(800-900)$ for
many different reasons which we shall discuss shortly.

Given the new CLEO measurement~\cite{cleo} of the 
$a_{1}(1230)\rightarrow\sigma\pi$
branching ratio based on $\tau\rightarrow\nu3\pi$ decay of
$\mathrm{BR}(a_{1}\rightarrow\sigma\pi)=(16\pm4)\%$, the average
PDG value of~\cite{pdg} $\Gamma(a_{1})\sim425$ MeV then suggests
a substantial partial width of size
\begin{equation}
  \label{eq:cleo}
  \Gamma_{\mathrm{CLEO}} (a_{1}\rightarrow\sigma\pi) \sim 
  (0.16)(425 \ \mathrm{MeV}) = 68 \pm 33 \ \mathrm{MeV}\,.
\end{equation}
This was anticipated a decade ago by Weinberg~\cite{weinb}, using
mended chiral symmetry (MCS) to predict
\begin{equation}
  \label{eq:weinb}
   \Gamma_{\mathrm{MCS}} (a_{1}\rightarrow\sigma\pi) = 
   2^{-3/2}\Gamma_{\rho} \approx 53 \ \mathrm{MeV} \, .
\end{equation}
Moreover, assuming chiral symmetry, the needed coupling is related
to $g_{a_{1}\sigma\pi}=g_{\rho\pi\pi}\approx6$, the latter found from 
$\Gamma_{\rho}\approx151$ MeV.
Invoking the PDG $\sigma$ mass of $\sim550$ MeV~\cite{pdg,tornquist}
(giving $q_{\mathrm{CM}}\approx480$ MeV), one anticipates the width
\begin{equation}\label{eq:ours}
  \Gamma (a_{1}\rightarrow\sigma\pi) = \frac{1}{3}
  \left( g^{2}_{a_{1}\sigma\pi}/4\pi \right) 
  \frac{q^{3}_{\mathrm{CM}}}{m^{2}_{a_{1}}} 
  \approx 70 \ \mathrm{MeV} \, .
\end{equation}

Considering the compatible (nonvanishing) 
$\Gamma_{a_{1}\rightarrow\sigma\pi}$
widths in Eqs.~(\ref{eq:cleo}--\ref{eq:ours}) above, one might
question (as Weinberg did in reference~\cite{weinb}) why the PDG
listed the much smaller value 
BR$(a_{1}\rightarrow\pi(\pi\pi)_{\mathrm{swave}})<0.7\%$ in the
$1980$s or the essentially vanishing width
\begin{equation}
  \label{eq:pmone}
  \Gamma(a_{1}\rightarrow\pi(\pi\pi)_{\mathrm{swave}}) = 
  1 \pm 1 \ \mathrm{MeV}
\end{equation}
in the $1990$s.

\section{Vanishing Soft Pion Theorem (SPT)}

To resolve this apparent contradiction, we note that there are
in fact \emph{two} Feynman graphs to consider for
$a_{1}\rightarrow\pi(\pi\pi)_{\mathrm{swave}}$ decay, the ``box''
quark graph of Fig.~$1$a and the quark ``triangle'' graph of Fig.~$1$b
(for nonstrange $u$ and $d$ quarks). 
In the soft pion limit for one soft pion in the 
$(\pi\pi)_{\mathrm{swave}}$ doublet (but not the pion outside the
$(\pi\pi)_{\mathrm{swave}}$ doublet),
there is a vanishing SPT~\cite{ivanov,ivanov1}, cancelling the box graph in
Fig.~1a against the triangle graph Fig.~1b in the chiral soft pion limit.

Such a cancellation stems from the Dirac matrix 
\emph{identity}\footnote{Equation~(\ref{eq:identity}) reduces to 
$2m\gamma_{5} = 2m\gamma_{5}$ when multiplying both sides of~(\ref{eq:identity}) 
on the lhs and rhs by $(\gamma \cdot p - m)$.}
\begin{equation}
  \label{eq:identity}
  \frac{1}{\gamma\cdot p - m} 2m\gamma_{5} \frac{1}{\gamma\cdot p - m}
  \equiv -\gamma_{5} \frac{1}{\gamma\cdot p - m} - 
         \frac{1}{\gamma\cdot p - m} \gamma_{5} \, .
\end{equation}
We apply~(\ref{eq:identity}) together with the pseudoscalar pion quark (chiral) 
Goldberger--Treiman coupling $g_{\pi qq} = m / f_{\pi}$ for 
$f_{\pi} \approx 93$ MeV.
This SPT for $p_{\pi}\rightarrow 0$ applied to graphs of Figs.~1--4 results
in 

a) $a_{1}\rightarrow\pi(\pi\pi)_{\mathrm{swave}}$:

The box graph of Fig.~1a and Eq.~(\ref{eq:identity}) gives the amplitude as
$p_{\pi}\rightarrow 0$,
\begin{equation}
  \label{eq:two}
    M^{box}_{a_{1}\rightarrow 3\pi} \rightarrow - \frac{1}{f_{\pi}} 
    M(a_{1}\rightarrow \sigma\pi) \, .
\end{equation}
But the additional $\sigma$ pole quark triangle graph of Fig.~1b is
\begin{equation}
  \label{eq:three}
    M^{tri}_{a_{1}\rightarrow 3\pi} = \frac{1}{f_{\pi}} 
    M(a_{1}\rightarrow \sigma\pi) \, ,
\end{equation}
because $2g_{\sigma\pi\pi}=(m^{2}_{\sigma}-m^{2}_{\pi})/f_{\pi}$ in the
linear $\sigma$ model (L$\sigma$M).
Thus the sum of~(\ref{eq:two}) and~(\ref{eq:three}) vanishes in the soft pion
limit~\cite{ivanov,ivanov1}
\begin{equation}
  \label{eq:four}
    M_{a_{1}\rightarrow 3\pi}|_{total} =
    M^{box}_{a_{1}\rightarrow 3\pi} +
    M^{tri}_{a_{1}\rightarrow 3\pi} \rightarrow 0 \, ,
\end{equation}
compatible with data~\cite{pdg}: 
$\Gamma(a_{1}\rightarrow\pi(\pi\pi)_{\mathrm{swave}})=1\pm1$ MeV.

b) $\gamma\gamma\rightarrow2\pi^{0}|_{s=m^{2}_{\sigma}}$:

Again using pseudoscalar pion-quark couplings, it was predicted~\cite{kaloshin} 
five years before data appeared that this $\gamma\gamma\rightarrow2\pi^{0}$  
cross section should fall to about $10$ nbarns in the $700$ MeV region.
Equivalently, using the SPT theorem stemming from Eq.~(\ref{eq:identity}), we
predict the amplitude due to the quark box plus quark triangle graphs of
Fig.~2
\begin{equation}
  \label{eq:five}
    \langle \pi^{0}\pi^{0}|\gamma\gamma \rangle \rightarrow
    \left[ - \frac{i}{f_{\pi}}\langle\sigma|\gamma\gamma\rangle +
    \frac{i}{f_{\pi}}\langle\sigma|\gamma\gamma\rangle \right] \rightarrow 0 \, ,
\end{equation}
as $s\rightarrow m^2_{\sigma}(700)$~\cite{ivanov1}.
This picture was supported by recent Crystal Ball data~\cite{crystalball}.

c) $\pi^{-}p\rightarrow\pi^{-}\pi^{+}n$:

The SPT stemming from Eq.~(\ref{eq:identity}) also suggests that the sum of the 
two $\pi^{+}$ peripheral--dominated  $\pi^{-}p\rightarrow\pi^{-}\pi^{+}n$ amplitudes
of Figs.~3 vanishes:
\begin{equation}
  \label{eq:six}
    M_{\pi^{-}p\rightarrow\pi^{-}\pi^{+}n}|_{per} \propto 
    \left[ M^{box}_{\pi\pi} + M^{tri}_{\pi\pi} \right] \rightarrow 0 \, .
\end{equation}
This ``chirally--eaten'' $\sigma(600-700)$ in Figs.~1b, 2b, 3b indeed did 
not appear in PDG tables prior to 1996, just as the SPT mandates.
In fact the $\sigma(600-700)$ does not appear in recent Crystal Ball 
$\pi^{-}p\rightarrow\pi^{0}\pi^{0}n$ studies either~\cite{nefkens}.

d) $K^{-}p\rightarrow K^{-}\pi^{+}n$:

Finally the SPT due to Eq.~(\ref{eq:identity}) requires the sum of the two 
$\pi^{+}$ peripheral--dominated  $K^{-}p\rightarrow K^{-}\pi^{+}n$ amplitudes
of Figs.~4 to vanish,
\begin{equation}
  \label{eq:seven}
    M_{K^{-}p\rightarrow K^{-}\pi^{+}n}|_{per} \propto 
    \left[ M^{box}_{K\pi} + M^{tri}_{K\pi} \right] \rightarrow 0 \, ,
\end{equation}
shielding this ground state $\kappa^{0}(800-900)$ scalar in Fig.~4b.
Instead the $K^{\ast}(1430)$ (excited state) scalar resonance clearly 
appears in LASS data~\cite{lass}; this $K^{\ast}(1430)$ not being eaten
means it also is not a true ground state \mbox{scalar} obeying the SPT.
An analogous disappearance of the ground state $\kappa(800-900)$
scalar occurs for the peripheral-dominated processes
$K^{-}p\rightarrow \pi^{-}\pi^{+}\Lambda, \bar{K}K\Lambda$.

None of the above four SPT processes depicted in Figs.~1--4 have been 
used by the experimentalists to observe such scalar mesons. 
Instead they study processes avoiding these four SPTs, e.g. 
J/$\psi\rightarrow \omega\pi\pi$ to isolate the $\sigma(500)$ resonance 
`bump'.
In effect, the above s--wave SPTs (with quark boxes cancelling
quark triangle graphs in the soft pion limit) chirally `eat'
the ground state $\sigma(600-700)$ and $\kappa(800-900)$ scalar
mesons, justifying in 
part\footnote{Two other reasons for suppressing these scalars
are: (1) they are low mass and broad, sometimes at the edge of the 
phase space and (2) they are usually swamped by the nearby vectors 
$\rho(770)$ or $\omega(783)$ and $K^{\ast}(895)$, respectively.}
why these scalar mesons have been so 
difficult to isolate and identify in the past.

With hindsight, the L$\sigma$M dynamically generates ground state 
$\sigma(650)$ and $\kappa(850)$ scalars via (one-loop-order)
tadpole graphs~\cite{nontriv}.
Even though these tadpoles can be suppressed by working in the infinite 
momentum frame \mbox{(IMF)}~\cite{imf}, SU(6) mass formulae (requiring squared 
masses) then kinematically favor~\cite{scadron} the (ground state) 
$\sigma(650)$ and $\kappa(820)$.
This is another way (besides e.g. J/$\psi\rightarrow \omega\pi\pi$) to 
circumvent the four SPTs discussed in this section.

\section{Quark Loops versus Meson Loops}

In most effective chiral field theories (such as the L$\sigma$M), one usually 
computes consistently either quark loops alone or meson loops alone for 
a given process.
Sometimes one must add together quark and meson loops~\cite{nontriv}.
Chiral symmetry and the SPT discussed in Sec. II actually help to put 
order in this morass of quarks and meson loops.

Specifically for PVV transitions, the anomaly~\cite{abj} or simply the
vanishing of e.g. a meson $\pi\pi\pi$ vertex, etc. leads directly to a 
`quark loops alone' theory~\cite{pvv}, such as for $\pi^{0}\rightarrow 2\gamma$.
However for SVV transitions, it turns out that \emph{only} meson loop 
graphs contribute.
This SVV `meson loops alone' theorem also is a direct consequence of the soft
pion theorem (SPT) proved in refs.~\cite{ivanov,ivanov1} and reviewed 
in Sec. II above.
Specifically we study $\gamma\gamma\rightarrow\pi^{0}\pi^{0}$ with one
of the pions soft.
Again the quark box plus quark triangle graphs of Figs.~2 add up to 
zero in the soft pion limit.
Turning Fig.~2b around, if $\sigma$ (as a $2\pi$ resonance) decays
to $2\gamma$, this SPT eats up the needed quark triangle due to the
quark box.
This leaves only the meson triangle 
$\sigma\rightarrow K^{+}K^{-} \rightarrow 2 \gamma$ dominating
SVV decay $\sigma\rightarrow\gamma\gamma$.

A more practical example of this theorem is for
$a_{0}(983)\rightarrow 2\gamma$ decay.
First we consider the inverse process $\gamma\gamma\rightarrow\eta\pi$,
with the $\eta\pi$ final state forming an $a_{0}(983)$ resonance
$\gamma\gamma\rightarrow a_{0} \rightarrow \eta\pi$. 
So we should begin by first considering the quark box graph for
$\gamma\gamma\rightarrow a_{0}$ followed by
$a_{0}\rightarrow\eta\pi$.
Again these quark box plus triangle graphs vanish in the soft pion 
limit  by the SPT  of Sec. II.
All that remains are the meson loop graphs for 
$a_{0}\rightarrow\gamma\gamma$ decay.

Here $a_{0}\rightarrow K^{+}K^{-} \rightarrow 2 \gamma$ and the charged kaon 
loop contributes to the $a_{0}\gamma\gamma$ covariant amplitude
\begin{equation}\label{eq:covampl}
  \langle 2\gamma | a_{0} \rangle = \mathrm{M} \varepsilon_{\mu}(k') 
  \varepsilon_{\nu}(k) (g^{\mu\nu}k'\cdot k - k'^{\mu}k^{\nu}) \,
\end{equation}
where, according to ref.~\cite{deakin}, the effective amplitude $\mathrm{M}$
is given by
\begin{equation}\label{eq:M}
  | \mathrm{M}_{\mathrm{K-loop}} | = \frac{2g'\alpha}{\pi m^{2}_{a_{0}}}
  \left[ - \frac{1}{2} + \xi I(\xi) \right] \, ,
\end{equation}
with $\xi=m^{2}_{K^{+}}/m^{2}_{a_{0}}=0.2520>1/4$. 
Then the loop integral becomes
\begin{equation}\label{eq:int_xi}
  I(\xi) = \int_{0}^{1} \! dy \, y \int_{0}^{1} \! dx 
  \left[ \xi - xy(1-y) \right]^{-1} = 2 
    \left[ \arcsin\sqrt{1/4\xi} \right]^{2} \approx 4.39 \, .
\end{equation}
Also the L$\sigma$M $a_{0}KK$ coupling ($g'$) is~\cite{deakin,su3}
\begin{equation}\label{eq:gprime}
  g' = (m^{2}_{a_{0}} - m^{2}_{K})/2f_{K} \approx 3.18 \ \mathrm{GeV} \, ,
\end{equation}
so that the $a_{0}\gamma\gamma$ amplitude in Eq.~(\ref{eq:M}) is
approximately
\begin{equation}\label{eq:Mvalue}
  | \mathrm{M}_{\mathrm{K-loop}} | \approx 9.27 \times 10^{-3} \ \mathrm{GeV}^{-1} \, .
\end{equation}
This results in the decay width
\begin{equation}\label{eq:a0width}
  \Gamma ( a_{0} \rightarrow 2 \gamma ) = m^{3}_{a_{0}} | \mathrm{M}_{K} |^{2} /
  64 \pi \approx 0.406 \ \mathrm{keV} \, .
\end{equation}
The resonance $\kappa(900)$ contributes~\cite{deakin} $10\%$ of
Eq.~(\ref{eq:Mvalue}), reducing~(\ref{eq:a0width}) to
\begin{equation}\label{eq:a0width1}
  \Gamma ( a_{0} \rightarrow 2 \gamma ) 
  \approx 0.406 \ \mathrm{keV} (0.90)^{2}
  \approx 0.33 \ \mathrm{keV} \, .
\end{equation}
Assuming the $a_{0}$ width is ($100\%$) dominated by 
$a_{0}\rightarrow\eta\pi$, the PDG tables suggest
\begin{equation}\label{eq:a0widthPDG}
  \Gamma ( a_{0} \rightarrow 2 \gamma ) = 
  \left( 0.24^{+0.08}_{-0.07} \right) \ \mathrm{keV} \, .
\end{equation}

Another measured SVV decay is $f_{0}(980)\rightarrow\gamma\gamma$
with~\cite{pdg}
\begin{equation}\label{eq:f0widthPDG}
  \Gamma ( f_{0} \rightarrow 2 \gamma ) = 
  0.56 \pm 0.11 \  \mathrm{keV} \, .
\end{equation}
Here $\sigma-f_{0}$ mixing enters the amplitude analysis
with~\cite{su3,scadron84}
\begin{equation}\label{eq:mixing}
  |f_{0}\rangle = \sin \phi_{s} |\mathrm{NS}\rangle +
                  \cos \phi_{s} |\mathrm{S}\rangle  \, ,
\end{equation}
for $f_{0}(980)$ being mostly strange, with 
$\phi_{s}\approx 20^{\circ}$.
The nonstrange (NS) and strange (S) quark basis states are
respectively $|\mathrm{NS}\rangle = |\bar{u}u+\bar{d}d\rangle/\sqrt{2}$
and $|S\rangle=|\bar{s}s\rangle$ with singlet-octet angle 
$\theta_{s} = \phi_{s} - \arctan \sqrt{2}$.
The angle $\phi_{s}$ can be obtained from Eq.~(\ref{eq:mixing}) using
$\langle\sigma | f_{0} \rangle = 0 $ or 
$m^{2}_{\sigma_{s}} = m^{2}_{\sigma} \sin^{2} \phi_{s} + 
m^{2}_{f_{0}} \cos^{2} \phi_{s} $,
leading to~\cite{su3,scadron84}
\begin{equation}\label{eq:mixing_angle}
  \phi_{s} = \arcsin \left[ 
  \frac{m^{2}_{f_{0}} - m^{2}_{\sigma_{s}}}
       {m^{2}_{f_{0}} - m^{2}_{\sigma}} \right]^{1/2}\approx20^{\circ}
\end{equation}
for $m_{\sigma}\approx610$ MeV and 
$m_{\sigma_{s}}\approx 2m_{s} \approx 940$ MeV, with constituent
quark masses $m_{s} = ( m_{s} / \hat{m} ) \hat{m} \approx 470$ MeV,
and $\hat{m}\approx 325$ MeV, $m_{s}/\hat{m}\approx 1.45$.
Since $f_{0}(980)$ is mostly $\bar{s}s$ with 
$m_{f_{0}}\approx m_{a_{0}}$~\cite{su3}, we simply scale up the width
$\Gamma_{a_{0} \rightarrow \gamma\gamma} \approx 0.33$ keV in 
Eq.~(\ref{eq:a0width1}) by $2(\cos20^{\circ})^{2}$ from
Eq.~(\ref{eq:mixing}) (the $2$ due 
to~\cite{su3,scadron84} $g_{SKK}=1/\sqrt{2}$ whereas $g_{NSKK}=1/2$):
\begin{equation}\label{eq:f0width}
\Gamma (f_{0}\rightarrow\gamma\gamma) \approx 2 (\cos 20^{\circ})^{2}
(0.33\ \mathrm{keV})\approx 0.58 \ \mathrm{keV} \, ,
\end{equation}
again for a $f_{0} \rightarrow K^{+}K^{-} \rightarrow 2\gamma$ meson loop.

We observe that the predictions (\ref{eq:a0width1}) and (\ref{eq:f0width}) 
are in close agreement with the $a_{0},f_{0}\rightarrow 2\gamma$ measured 
decay rates in (\ref{eq:a0widthPDG}) and (\ref{eq:f0widthPDG}), respectively.

%
%
%
%
%
%
%

\section{Summary}

In Sec. I we gave one experimental and two theoretical reasons supporting the
somewhat broad width
$\Gamma (a_{1}\rightarrow\sigma\pi) \sim 65$ MeV.
The latter appears to contradict the complementary PDG result 
$\Gamma (a_{1}\rightarrow\pi(\pi\pi)_{\mathrm{swave}}) = 1 \pm 1$ MeV.
But in Sec. II we resolve this apparent contradiction, finding that 
\emph{both} quark box and quark triangle graphs contribute to the rate
$\Gamma (a_{1}\rightarrow\pi(\pi\pi)_{\mathrm{swave}})$, but the quark
box--triangle sum of these amplitudes \emph{vanishes} in the soft--pion
limit.
This SPT is also valid for $\sigma(\gamma\gamma\rightarrow\pi^{0}\pi^{0})$,
and peripheral decay rates 
$\Gamma_{\mathrm{per}} (\pi^{-} p \rightarrow \pi^{-}\pi^{+} n)$, 
$\Gamma_{\mathrm{per}} (K^{-} p \rightarrow K^{-}\pi^{+} n)$.
With hindsight, our quark loop chiral shielding SPTs in Sec. II parallel
the L$\sigma$M ``miraculous cancellation'' eating up the $\sigma$ pole in $\pi-\pi$ 
scattering ref.~\cite{fubini}, reducing the low energy amplitude
to Weinberg's well-known CA--PCAC result~\cite{weinb66}.
Finally in Sec. III we turn this SPT around.
Not only are pseudoscalar meson PVV decays controlled by quark loops
alone (as is well known e.g. for $\pi^{0}\rightarrow2\gamma$), but
scalar meson SVV decays are governed by meson loops alone.
We demonstrate how this latter SVV theorem works for
$a_{0}\rightarrow2\gamma$ and $f_{0}\rightarrow2\gamma$ decays.

Without invoking this SPT, there are physicists who do appreciate
the utility of a meson loop only scheme for SVV decays~\cite{lucio}.

\medskip
{\bf Acknowledgments}: The authors LB and MDS appreciate discussions 
with V.~Elias and partial support by the US DOE.
Authors YAB and ANI acknowledge discussions with V.~F.~Kosmach and
N.~I.~Troitskaya.


\newpage

\bigskip

\centerline{\bf Figure Captions.}

\noindent
Fig.~1: Quark $u$, $d$ box (a) and triangle (b) graphs contributing to 
$a_{1}\rightarrow\pi(\pi\pi)_{\mathrm{swave}}$.

\noindent
Fig.~2: Quark $u$, $d$ box (a) and triangle (b) graphs contributing to 
$\gamma\gamma\rightarrow\pi^{0}\pi^{0}$.

\noindent
Fig.~3: Peripheral--dominated quark $u$, $d$ box (a) and triangle (b) 
graphs contributing to $\pi^{-} p \rightarrow \pi^{-}\pi^{+} n$. 

\noindent
Fig.~4: Peripheral--dominated quark $u$, $d$ box (a) and triangle (b) 
graphs contributing to $K^{-} p \rightarrow K^{-} \pi^{+} n$.


\newpage

\bigskip

\begin{thebibliography}{99}

\bibitem{plethora} 
	E.~Beveren, G.~Rupp, Phys. Lett. {\bf B454}, 165 (1999), hep-ph/9902301;
	A.~V.~Anisovich, V.~V.~Anisovich, D.~V.~Bugg, V.~A.~Nikonov, Phys. Lett. {\bf B456},
	80 (1999), hep-ph/9903396;
	N.~A.~Tornqvist, ``Does the light and broad $\sigma(500)$ exist?'', hep-ph/9904346;
	T.~Hannah, Phys. Rev. {\bf D60}, 017502 (1999), hep-ph/9905236;
	M.~R.~Pennington, ``Riddle of the scalars: where is the $\sigma$?'', hep-ph/9905241;
	P.~Minkowski, W.~Ochs, ``The $J^{PC} = 0^{++}$ scalar meson nonet and glueball of 
	lowest mass'', hep-ph/9905250;
	S.~Ishida, M.~Ishida, ``Covariant Classification of $q\bar{q}$-Meson Systems and 
	Existence of New Scalar and Axial-Vector Mesons'', hep-ph/9905258;
	S.~Ishida, ``Controversies on and a Reasoning for Existence of the light 
	$\sigma$-particle'', hep-ph/9905260;
	M.~Ishida, S.~Ishida, T.~Ishida, K.~Takamatsu, T.~Tsuru, ``Observed Properties of 
	$\sigma$-particle'', hep-ph/9905261;
	T.~Kunihiro, ``Significance of the $\sigma$ Meson in Hadron Physics (QCD) and 
	Possible Experiments to Observe it'', hep-ph/9905262;
	T.~G.~Steele, F.~Shi, V.~Elias, ``QCD Sum-Rule Invisibility of the $\sigma$ Meson'',
	hep-ph/9905303.

\bibitem{cleo}
	D.~Asner et al., CLEO Collaboration, ``Hadronic Structure in the Decay 
	$\tau^{-}\rightarrow\nu_{\tau}\pi^{-}\pi^{0}\pi^{0}$ and the Sign of the 
	Tau Neutrino Helicity'', hep-ex/9902022.

\bibitem{pdg} 
	Particle Data Group, C.~Caso et al., Eur. Phys. J. {\bf C3}, 1 (1998).

\bibitem{weinb}
	S.~Weinberg, Phys. Rev. Lett. {\bf 65}, 1177 (1990).	

\bibitem{tornquist} 
	N.~A.~Tornquist and M.~Roos, Phys. Rev. Lett. {\bf 76}, 1575 (1996); 
	M.~Harada, F.~Sannino and J.~Schechter, Phys. Rev. {\bf D54}, 1991 (1996);
	S.~Ishida et al., Prog. Theor. Phys. {\bf 95}, 745 (1996); 
	S.~Ishida et al., ibid {\bf 98}, 621 (1997); for the theoretical linear 
	$\sigma$ model version, see R.~Delbourgo and M.~D.~Scadron, Mod. Phys. Lett.
	{\bf A10}, 251 (1995), hep-ph/9910242.

\bibitem{ivanov} 
	A.~N.~Ivanov, M.~Nagy, and M.~D.~Scadron, Phys. Lett. {\bf B273}, 137 (1991).

\bibitem{ivanov1} 
	A.~N.~Ivanov, M.~Nagy, and N.~I.~Troitskaya, Mod. Phys. Lett. {\bf A7}, 1997 (1992); 
	M.~D.~Scadron, Phys. At. Nucl. {\bf 56}, 1595 (1993).

\bibitem{kaloshin} 
	A.~E.~Kaloshin and V.~V.~Serebryakov, Zeit. Phys. {\bf C32}, 279 (1986).

\bibitem{crystalball} 
	H.~Marsiske et al., Crystal Ball Collaboration, Phys. Rev. D{\bf 41}, 3324 (1990); 
	J.~Bienlein, 4-th Int. Workshop on $\gamma\gamma$ collisions, La Jolla (1992).

\bibitem{nefkens} 
	B.~M.~K.~Nefkens and A.~B.~Starostin, ``Hadron
   	Physics with the Crystal Ball'' in $\pi N$ Newsletter, No. 15, December
   	1999, pp. 78--83 (see e.g. Fig.~2). This is the neutral pion analogue of 
	the charged pion peripheral process $\pi^{-} p \rightarrow \pi^{-} \pi^{+} n$ 
	considered in the text.

\bibitem{lass} 
	D.~Aston et al., LASS Collaboration, Nucl. Phys. B296, 493 (1998).

\bibitem{nontriv}
	Delbourgo--Scadron in ref.~\cite{tornquist}; also see R.~Delbourgo, A.~Rawlinson,
	and M.~D.~Scadron, Mod. Phys. Lett. {\bf A13}, 1893 (1998), hep-ph/9807505; 
	L.~Babukhadia and M.~D.~Scadron, Eur. Phys. J. 
	{\bf C8}, 527 (1999), hep-ph/9812424. Adding quark loops to meson loops might
	appear confusing in the large $\mathrm{N}_{c}$ limit. In fact the above references
	show that in the quark level L$\sigma$M, the Lee L$\sigma$M condition (the sum of
	the quark plus meson tadpoles must vanish), leads to
	$\mathrm{N}_{c}(2m_{q})^{4}=3m^{4}_{\sigma}$ together with the chiral NJL relation
	$m_{\sigma} = 2 m_{q}$, requiring $\mathrm{N}_{c}=3$ and not
	$\mathrm{N}_{c}\rightarrow\infty$.
	Also see B.~W.~Lee, \emph{Chiral Dynamics} (Gordon and Breach, 1972), p.~12.

\bibitem{imf} 
	F.~Fubini and G.~Furlan, Physics 1, 229 (1965).

\bibitem{scadron} 
	M.~D.~Scadron, Phys. Rev. D{\bf 26}, 239 (1982); Mod. Phys. Lett. A7, 669 (1992).
	Also see the nonrelativistic unitarized meson model scheme of 
        E.~van~Beveren et al., Zeit. Phys. {\bf C30}, 615 (1986).

\bibitem{abj}
	The AVV anomaly triangle was first considered by S.~L.~Adler, Phys. Rev. 
	{\bf 177}, 2426 (1969); J.~S.~Bell and R.~Jackiw, Nuovo Cim. {\bf 60}, 47 (1969),
	or see J.~Steinberger, Phys. Rev. {\bf 75}, 651; {\bf 76}, 790 (1949).

\bibitem{pvv}
	R.~Delbourgo, D.~Liu and M.~D.~Scadron, Int. Journ. Mod. Phys. {\bf A14}, 4331 (1999),
	hep-ph/9905501.

\bibitem{deakin}
	A.~S.~Deakin, V.~Elias, D.~G.~C. McKeon, M.~D.~Scadron and A.~Bramon, Mod. 
	Phys. Lett. {\bf A9}, 2381 (1994).

\bibitem{su3}
	R.~Delbourgo and M.~D.~Scadron, Int. Journ. Mod. Phys. {\bf A13}, 657 (1998),
	hep-ph/9807504; R.~Delbourgo, D.~Liu and M.~D.~Scadron, Phys. Lett. {\bf B446},
	332 (1999).

\bibitem{scadron84}
	M.~D.~Scadron, Phys. Rev. {\bf D29}, 1375 (1984).

\bibitem{fubini}
	V.~deAlfaro, S.~Fubini, G.~Furlan and C.~Rossetti, ``Currents in Hadron
	Physics'', (North Holland $1973$). See pp $324$--$327$. Also see
	M.~D.~Scadron, Eur. Phys. J. {\bf C6}, $141$ ($1999$), hep-ph/$9710317$.

\bibitem{weinb66}
	S.~Weinberg, Phys. Rev. Lett. {\bf 17}, $616$ ($1966$).	

\bibitem{lucio}
	See e.g. J.~Lucio and M.~Napsuciale, hep-ph/9903234.

%
%
%
%
%

\end{thebibliography}
\end{document}